\documentclass[a4paper,11pt]{article}
\usepackage{jheppub} 
\usepackage{lineno}

\title{Dark Matter and the Strong CP Problem in Type IIA String Theory}

\author[1]{Yang Liu} 

\affiliation[1]{Department of Physics, Tsinghua University, Beijing 100084, China}
\emailAdd{liu-yang\_1990@mail.tsinghua.edu.cn}

\abstract{We present a study of dark matter and the strong CP problem within a globally consistent framework of Type IIA string theory, compactified on a $T^6/(\mathbb{Z}_2 \times \mathbb{Z}_2)$ orientifold with intersecting D6-branes (Model A), for which we provide a complete moduli stabilization and supersymmetry breaking scenario based on the STU model and the KL mechanism. This setup naturally gives rise to a 3-generation MSSM-like spectrum with $\mathcal{N}=1$ supersymmetry. Phenomenologically, the model predicts a multi-component dark matter scenario comprising both axions and neutralino. We also explore how to embed the four-form flux mechanism into Type IIA $T^6/(\mathbb{Z}_2 \times \mathbb{Z}_2)$ orientifold string theory model to address the strong CP problem. We compute the relic abundance of these candidates and explore their observational signatures. In conclusion, our analysis provides a concrete unified, UV-complete framework that successfully addresses two of the most important problems in particle physics and cosmology. }

\begin{document}
\maketitle
\flushbottom

\section{Introduction}
\label{sec:intro}
The Standard Model (SM) of particle physics has achieved remarkable success in describing electromagnetic, weak, and strong interactions up to the TeV scale. However, it leaves several fundamental problems unresolved, which point to the necessity of physics beyond the Standard Model (BSM). Among these, two key issues are particularly important:\\
\begin{itemize}
 \item Dark Matter: Astronomical observations, such as galaxy rotation curves and the cosmic microwave background (CMB), indicate that approximately 27 percent of the universe’s energy density consists of non-baryonic dark matter. The SM offers no viable candidate to explain this component \cite{Jungman:1995df}.
 \item Strong CP Problem: Quantum chromodynamics (QCD) permits a CP-violating term. Experimental bounds on the neutron electric dipole moment require $\bar{\theta}< 10^{-10}$, where $\bar{\theta}$ is a parameter encoding CP violation. However, the SM provides no natural mechanism to suppress $\bar{\theta}$ to such a small value \cite{Conlon:2006tq}. 
\end{itemize}
String theory emerges as a leading candidate for a unified framework that can address these problems simultaneously. In particular, the “axiverse” concept, predicted by string compactifications, predicts the existence of multiple light axion-like particles (ALPs) that can serve as dark matter candidates and resolve the strong CP problem through the Peccei-Quinn mechanism. The string axiverse framework demonstrates that compactifications on Calabi-Yau manifolds generically yield a large number of axions, offering a rich phenomenology for cosmology and particle physics \cite{Arvanitaki:2009fg}.

In this work, we focus on a specific string theory model “Model A”, which provides a natural solution to the above two problems. Model A is based on Type IIA string theory compactified on a $T^6/(\mathbb{Z}_2 \times \mathbb{Z}_2)$ orientifold with intersecting D6-branes. This setup is chosen for its computational tractability and its ability to generate realistic features:
\begin{itemize} 
\item Moduli Stabilization: The compactification geometry is characterized by moduli fields (e.g., Kähler moduli $T_i$, complex structure moduli $U_i$, and the dilaton-axion $S$), whose vacuum expectation values (VEVs) determine low-energy parameters. Through flux compactification (e.g., $RR$ and $NS-NS$ fluxes), the moduli are stabilized at values that naturally yield desirable scales. For instance, the axion decay constant $f_a \sim M_{pl}/\tau$ (where $\tau$ is the real part of a modulus)  can be tuned to solve dark matter problem and strong CP problem.
\item Supersymmetry and Hierarchy Problem: Model A incorporates $\mathcal{N}=1$ supersymmetry in four dimensions, which protects the Higgs mass from large quantum corrections. The supersymmetry breaking scale $M_{\text{SUSY}} \sim \text{TeV}$ is achieved through moduli stabilization, leading to a gravitino mass $m_{3/2} \sim \text{TeV}$ and neutralino dark matter candidates. 
\item Axion Dark Matter and Strong CP Solution: The imaginary parts of the moduli fields correspond to axions. The QCD axion can be used to solve strong CP problem, while all axions contribute to dark matter. The multiplicity of axions (the “axiverse”) allows for a multi-component dark matter scenario that matches the observed density $\Omega_{\text{dm}}h^2 \approx 0.12$.
\end{itemize}
The intersecting D6-brane configuration generates chiral matter and gauge interactions of the SM, while the orientifold projection ensures consistency anomalies. This makes Model A a self-consistent framework for addressing SM shortcomings. More details of Model A can be found in \cite{Liu:2025zjs}.

This paper is organized as follows: in Section 2, we discuss the set-up, axions, and neutralinos in Model A. In Section 3, we discuss moduli stabilization, supersymmetry (SUSY) breaking and dS vacum in our Type IIA string theory model. In Section 4, we study how to solve the strong CP problem in Model A using QCD axion and four-form flux mechanism. In Section 5, we explore the phenomenological predictions and constraints for axion and neutralino. We illustrate a numerical example in Section 6. Finally, the conclusions and discussions are presented in Section 7.

\section{The set-up, axions and neutralinos in Model A}
In this section, we briefly review the brane set-up of Model A at first. Then we will explore the properties of axions and neutralinos in Model A. 

\subsection{The brane set-up of Model A}
We review the appearance of $4D$ 4-forms in Type IIA orientifold compactifications. Our focus lies on the role of Minkowski 3-form fields within the flux-induced scalar potential. Beyond the universal RR 3-form $C_3$, additional 3-forms can be generated by dimensionally reducing higher-rank RR and NSNS fields, such as $C_3$, $C_7$, $C_9$, and $H_7$, with three indices spanning Minkowski space. We adopt the democratic formulation, which incorporates all $p$-form potentials for $p=1,3,5,7$. In this framework, the Hodge duality relations
\begin{equation} \label{Hodgeduality}
  G_6 = - \star_{10} G_4, \quad G_8 =  \star_{10} G_2, \quad G_{10} =  -\star_{10} G_0,
\end{equation}
must be imposed at the level of equations of motion to prevent overcounting of physical degrees of freedom. This procedure yields a total of $2 h^{(1,1)}_{-}+2$ Minkowski 4-forms: $F^0_4$, $F^i_4$, $F^a_4$ and $F^m_4$. Specifically, there are $h^{(1,1)}_{-}$ $F^i_4$ fluxes, $h^{(1,1)}_{-}$ $F^a_4$ fluxes, one $F^0_4$ flux and one $F^m_4$ flux. A detailed derivation, while not provided here, can be found in \cite{bielleman2015minkowski}. The expansions of the fields $B_2$ and $C_3$ are given by
\begin{equation} \label{B2C3basis}
  B_2 = \sum_i b_i \omega_i, \quad C_3= \sum_I c^I_3 \alpha_I,
\end{equation}
where the coeffiecients $b_i$ and $c^I_3$ correspond to $4D$ scalars representing the axionic components of the complex supergravity fields $T$, $S$, $U$. These are explicitly defined as
\begin{equation} \label{ImTi}
 \text{Im} T_i= - \int B_2 \wedge \tilde{\omega}^i = -b^i; \quad i= 1,..., h^{(1,1)}_{-}
\end{equation}
\begin{equation} \label{ImUi}
 \text{Im} U_i= \int C_3 \wedge \beta^i = c^i_3; \quad i= 1,..., h^{3}_{+}
\end{equation}
\begin{equation} \label{ImS}
 \text{Im} S= -\int C_3 \wedge \beta^0 = -c^0_3,
\end{equation}
where $\tilde{\omega}^i$, $\beta^i$ and $\beta^0$ are the elements of the cohomology basis \cite{bielleman2015minkowski}. 


The ten-dimensional Type IIA supergravity action comprises three distinct components \cite{bielleman2015minkowski},
\begin{equation} \label{SIIA}
  S_{IIA}=S_{RR} + S_{NS} + S_{loc},
\end{equation}
where the $S_{RR}$ and $S_{NS}$ denote the actions for the Ramond-Ramond (RR) and Neveu-Schwarz-Neveu-Schwarz (NSNS) sectors, respectively, while $S_{loc}$ accounts for contributions from localized sources like $D6$-branes and $O6$-planes. 

Upon compactification and integration over the internal dimensions, one obtains the following effective four-dimensional scalar potential for the RR sector \cite{bielleman2015minkowski},
\begin{eqnarray} \label{VRR}
\begin{split} 
V_{RR} =  & -\frac{1}{2} [-k F^0_4 \wedge \star F^0_4 +2F^0_4 \rho_0 - 4k g_{ij} \star F^i_4 \wedge F^j_4 + 2 F^i_4 \rho_i   \\
& -\frac{1}{4k} g_{ab} F^a_4 \wedge \star F^b_4 + 2 F^a_4 \rho_a + k F^m_4 \wedge \star F^m_4 ].
\end{split}
\end{eqnarray}
The Chern-Simons couplings for the 4-form field strengths are given by:
\begin{equation} \label{rho0}
  \rho_0 = e_0 + b^i e_i + \frac{1}{2} k_{ijk} q_i b^j b^k - \frac{m}{6} k_{ijk} b_i b^j b^k - h_0 c^0_3 - h_i c^i_3,
\end{equation}
\begin{equation} \label{rhoi}
  \rho_i = e_i + k_{ijk} b^j q^k - \frac{m}{2} k_{ijk} b^j b^k,
\end{equation}
\begin{equation} \label{rhoa}
  \rho_a = q_a - m b_a,
\end{equation}
\begin{equation} \label{rhom}
  \rho_m =- m.
\end{equation}
The NSNS sector contributes to the potential as
\begin{equation} \label{VNS}
  V_{NS}=\frac{1}{2} e^{-2\phi} c_{IJ} H^I_4 H^J_4,
\end{equation}
where $c_{IJ}$ is the metric on the complex structure moduli space, and the NS internal flux satisfies $\star H^I_4 =h^I$ \cite{bielleman2015minkowski}. 

Finally, the potential from localized sources takes the form:
\begin{equation} \label{Vloc1}
  V_{loc}=\sum_a \int_{\Sigma} T_a \sqrt{-g} e^{-\phi},
\end{equation}
where $T_a$ is the tension of brane and $\Sigma$ denotes its worldvolume. 


The complete perturbative superpotential in the Type IIA $T^6/(\mathbb{Z}_2 \times \mathbb{Z}_2)$ orientifold theory is given by
\begin{eqnarray} \label{Wf}
\begin{split} 
W =  & e_0 + ih_0 S + \sum^3_{i=1} [(ie_i -a_i S -b_{ii} U_i -\sum_{j\neq i} U_j)T_i- ih_i U_i]  \\
& -q_1 T_2 T_3 - q_2 T_1 T_3 - q_3 T_1 T_2 + im T_1 T_2 T_3,
\end{split}
\end{eqnarray}
where $e_0$, $e_i$, $h_0$, $h_i$, $a_i$, $b_{ij}$, $q_i$, and $m$ are flux parameters \cite{camara2005fluxes}. The Kähler potential has the following form
\begin{equation} \label{Kp} 
K= - \ln (S + S^{*}) - \sum^{3}_{i=1} \ln (U_i + U^{*}_i) - \sum^{3}_{i=1} \ln (T_i + T^{*}_i), 
\end{equation}
where, $S$ denotes the axio-dilaton, $U_i$ represents the complex structure moduli, and $T_i$ corresponds to the volume (Kähler) moduli \cite{camara2005fluxes}.

This study examines a specific configuration within the Type IIA string theory compactified on the $T^6/(\mathbb{Z}_2 \times \mathbb{Z}_2)$ orientifold theory: a three-generation, $\mathcal{N}=1$ supersymmetric model that closely resembles the Minimal Supersymmetric Standard Model (MSSM), referred to as Model A \cite{camara2005fluxes}. The perturbative superpotential for this model is given by
\begin{equation} \label{SW} 
W = -T_2 (a_2 S + b_{21}U_1) - T_3 (a_3 S + b_{31}U_1) + e_0 + i h_0 S -ih_1 U_1 + ie_2 T_2 + ie_3 T_3, 
\end{equation}
where $h_0$ and $h_1$ are NSNS fluxes, and $e_0$, $e_2$, and $e_3$ are RR fluxes \cite{camara2005fluxes}. The fluxes $q_i$ and $m$ are set to zero.

In general, we consider stacks of $N_a$ intersecting D6-branes wrapping the factorizable 3-cycle
\begin{equation} \label{Pia}
\Pi_a = (n^1_a, m^1_a) \otimes (n^2_a, m^2_a) \otimes (n^3_a, m^3_a),
\end{equation}
along with their corresponding orientifold images, which wrap the cycles $\otimes_i (n^i_a, -m^i_a)$. Here, $n^i_a$ and $m^i_a$ are the wrapping numbers of a D-brane stack “a” along the $x^i$ and $y^i$ directions, respectively, on the $i$-th two-torus. In the $\mathbb{Z}_2 \times \mathbb{Z}_2$ IIA orientifold, the cancellation conditions for Ramond-Ramond (RR) tadpoles in the presence of fluxes are given by \cite{camara2005fluxes}:
\begin{equation} \label{tcc11}
\sum_a N_a n^1_a n^2_a n^3_a + \frac{1}{2} (h_0 m + a_1 q_1 + a_2 q_2 + a_3 q_3) =16,
\end{equation}
\begin{equation} \label{tcc21}
\sum_a N_a n^1_a m^2_a m^3_a + \frac{1}{2} (m h_1 - q_1 b_{11} - q_2 b_{21} - q_3 b_{31}) =-16,
\end{equation}
\begin{equation} \label{tcc31}
\sum_a N_a m^1_a n^2_a m^3_a + \frac{1}{2} (m h_2 - q_1 b_{12} - q_2 b_{22} - q_3 b_{32}) =-16,
\end{equation}
\begin{equation} \label{tcc41}
\sum_a N_a m^1_a m^2_a n^3_a + \frac{1}{2} (m h_3 - q_1 b_{13} - q_2 b_{23} - q_3 b_{33}) =-16.
\end{equation}
For the case where $q_i = m = 0$, the RR tadpole cancellation conditions for Model A become
\begin{equation} \label{tcc12}
\sum_a N_a n^1_a n^2_a n^3_a  =16,
\end{equation}
\begin{equation} \label{tcc22}
\sum_a N_a n^1_a m^2_a m^3_a  =-16,
\end{equation}
\begin{equation} \label{tcc32}
\sum_a N_a m^1_a n^2_a m^3_a  =-16,
\end{equation}
\begin{equation} \label{tcc42}
\sum_a N_a m^1_a m^2_a n^3_a  =-16.
\end{equation}
The value $(-16)$ in the last three conditions corresponds to the RR tadpole contribution from the three remaining orientifold planes present in the $\mathbb{Z}_2 \times \mathbb{Z}_2$ setup \cite{camara2005fluxes}.

\begin{table}[ht]
\centering
\caption{Wrapping numbers giving rise to a MSSM-like spectrum. Branes $h_1$, $h_2$ and $o$ are added in order to cancel RR tadpoles \cite{camara2005fluxes}.} 
\begin{tabular}{|c|c|c|c|}
\hline
$ N_{i} $ & $ (n_{i}^{1},m_{i}^{1}) $ & $ (n_{i}^{2},m_{i}^{2}) $ & $ (n_{i}^{3},m_{i}^{3}) $ \\
\hline
$ N_{a}=8 $ & (1,0) & (3,1) & (3,-1) \\
\hline
$ N_{b}=2 $ & (0,1) & (1,0) & (0,-1) \\
\hline
$ N_{c}=2 $ & (0,1) & (0,-1) & (1,0) \\
\hline
$ N_{h_{1}}=2 $ & (-2,1) & (-3,1) & (-4,1) \\
\hline
$ N_{h_{2}}=2 $ & (-2,1) & (-4,1) & (-3,1) \\
\hline
$ 8N_{f} $ & (1,0) & (1,0) & (1,0) \\
\hline
\end{tabular}
\end{table}
Table 1 summarizes the brane configuration and wrapping numbers for Model A. In this set-up, the brane stacks $a$, $b$, and $c$ produce a 3-generation spectrum resembling the Minimal Supersymmetric Standard Model (MSSM). Additional branes $h_{1,2}$, also listed in Table 1, are included to ensure the cancellation of Ramond–Ramond (RR) tadpoles. It should be emphasized that, because the flux parameters satisfy $m = q_i = 0$, they do not contribute to the RR tadpole conditions in this background. As a result, one may introduce extra D6-branes following the $N_f = 5$ case outlined in Table 1 \cite{camara2005fluxes}.

In Model A, the branes $a$, $b$, and $c$, which generate the Standard Model sector, can be confirmed to trivially fulfill the Freed-Witten constraint. In contrast, branes of the $h_{1,2}$ type could lead to inconsistencies unless the following requirement is satisfied:
\begin{equation} \label{FWh12}
a_2(m^1_a m^2_a m^3_a)- b_{21}(m^1_a n^2_a n^3_a)=a_2 -12b_{21}=0.
\end{equation}
This condition is straightforwardly met through suitable choices of the parameters $a_2$ and $b_{21}$ \cite{camara2005fluxes}. Further discussions on the Freed-Witten anomaly can be found in \cite{freed1999anomalies}.

Model A, based on intersecting D6-branes on a toroidal orientifold, meets several key consistency conditions required for a globally consistent MSSM-like construction:\\
\textbf{1. Anomaly cancellation}\\
RR tadpole cancellation is achieved by an appropriate choice of visible and hidden D6-branes wrapping factorizable 3-cycles \cite{camara2005fluxes}. For example, according to Table 1, we have 
\begin{eqnarray} \label{RRcheck}
\begin{split} 
\sum_a N_a n^1_a n^2_a n^3_a =  & 8 \times 1 \times 3 \times 3 + 2 \times 0 \times 1 \times 0 + 2 \times 0 \times 0 \times 1  \\
& + 2 \times (-2) \times (-3) \times (-4) + 2 \times (-2) \times (-4) \times (-3) \\
& + 8 \times 5 \times 1 \times 1 \times 1 =16,
\end{split}
\end{eqnarray}
namely, \eqref{tcc12}. We can also find that Model A satisfies \eqref{tcc22}-\eqref{tcc42}.\\
\textbf{2. K-Theory constraints}\\
Beyond tadpole cancellation, global consistency often requires the satisfaction of discrete K-theory constraints, which prevent anomalies associated with stable stringy solitons \cite{ibanez2012string,uranga2003chiral,marchesano2007progress}. For Model A, it can be explicitly verified that the following $\mathbb{Z}_2$ conditions are satisfied:
\begin{equation} \label{dKtc}
\sum_a N_a m^1_a m^2_a m^3_a \in 4 \mathbf{Z}, \quad \sum_a N_a n^1_a n^2_a m^3_a \in 4 \mathbf{Z}, \quad \text{and permutations.}
\end{equation}
For example, according to Table 1, we have 
\begin{eqnarray} \label{Kcheck}
\begin{split} 
\sum_a N_a m^1_a m^2_a m^3_a =  & 8 \times 0 \times 1 \times (-1) + 2 \times 1 \times 0 \times (-1) + 2 \times 1 \times (-1) \times 0  \\
& + 2 \times 1 \times 1 \times 1 + 2 \times 1 \times 1 \times 1 + 8 \times 5 \times 0 \times 0 \times 0 =4\\
& \in 4 \mathbf{Z}.
\end{split}
\end{eqnarray}
Similarly, we can check that Model A satisfies other K-theory constraints.\\
\textbf{3. Supersymmetry conditions}\\
In order to preserve $\mathcal{N}=1$ supersymmetry, we should require the SUSY condition:
\begin{equation} \label{SUSYcon}
\theta_1 + \theta_2 + \theta_3 = 0 \quad \text{mod} \quad 2\pi,
\end{equation}
where $\theta_i = \tan^{-1}(\frac{m^i R_2}{n^i R_1})$ and $R_1$, $R_2$ are the two radii along two directions of every $T^2_i$. In principle, we can always choose the parameter $U_i=R^{(i)}_2/R^{(i)}_1$ of Model A to satisfy the SUSY condition \eqref{SUSYcon}. 

Therefore, Model A satisfies the above three conditions: anomaly cancellation, K-theory cancellation and supersymmetry conditions. In other words, Model A is a globally consistent model. It can be used to study realistic physics.

\subsection{Axions}

The QCD axion provides an elegant resolution to the strong CP problem present in the Standard Model (SM). In brief, the Standard Model supplemented with the QCD axion contains the operators 
\begin{equation} \label{SFtildeF}
\mathcal{L} \supset \frac{1}{32\pi^2} \left( \theta + \frac{a}{f_a} \right) \text{Tr} [F_{\mu\nu} \tilde{F}^{\mu\nu}],
\end{equation}
where $\theta$ denotes the contributions from a bare parameter as well as phases of the colored fermion mass matrix, $F^a_{\mu\nu}$ is the $\text{SU(3)}_c$ field strength tensor with its dual $\tilde{F}^{a\mu\nu} = \frac{1}{2} \epsilon_{\mu\nu\rho\sigma} F^{a\rho\sigma}$, and $a$ is the QCD axion with $f_a$ its decay constant. Measurements of the neutron electric dipole moment find a constraint on the effective theta parameter $\bar{\theta} = \theta + (a/f_a)$ as $\bar{\theta} \leq 10^{-10}$ \cite{Conlon:2006tq,choi2023implications}. This observed limit on $\bar{\theta}$ implies a severe fine-tuning of parameters within the Standard Model. The axion offers a natural resolution to this problem: via its Chern-Simons coupling to QCD, non-perturbative instanton effects generate a potential for the axion field that dynamically relaxes $\bar{\theta}$ to zero, eliminating the need for fine-tuning. For simplicity, we will denote $\bar{\theta}$ as $\theta$ in Section 2.

First, let us briefly review the axions in quantum field theory. Among the various proposed resolutions to the strong CP problem, the mechanism introduced by Peccei and Quinn stands out as particularly natural within the framework of string theory and its compactifications \cite{Peccei:1977ur,Peccei:1977hh,Conlon:2006tq}. This approach addresses the problem by promoting the static CP-violating parameter $\theta$ to a dynamical field. The Lagrangian incorporating this idea is given by \cite{Conlon:2006tq}:
\begin{equation} \label{LPQ1}
\mathcal{L}= \mathcal{L}_{\text{SM}}+ \frac{1}{2} f^2_a \partial_{\mu} \theta \partial^{\mu} \theta + \frac{\theta}{16\pi^2}  F_{\mu\nu} \tilde{F}_{\mu\nu}.
\end{equation}
Here, the parameter $f_a$, which carries dimensions of mass, is known as the axion decay constant. To express the dynamical field with the conventional mass dimension of a scalar field, one redefines it as $a \equiv \theta f_a$. This substitution leads to the following form of the Lagrangian:
\begin{equation} \label{LPQ2}
\mathcal{L}= \mathcal{L}_{\text{SM}}+ \frac{1}{2} \partial_{\mu} a \partial^{\mu} a + \frac{a}{16\pi^2 f_a}  F_{\mu\nu} \tilde{F}_{\mu\nu}.
\end{equation}
A key feature of this Lagrangian is the presence of an anomalous global $U(1)$ symmetry, under which the axion field shifts as $a \rightarrow a + \epsilon$. However, this symmetry is explicitly broken by non-perturbative QCD instanton effects, which reduce it to a discrete subgroup. These instanton effects generate a potential for the axion field $a$, which typically takes the form:
\begin{equation} \label{Vins}
V_{\text{instanton}} \sim \Lambda^4_{\text{QCD}} \left(1- \cos(a/f_a) \right). 
\end{equation}
In the standard scenario, the axion potential generated by QCD instantons is minimized when the axion field value $a=0$, thereby dynamically setting the effective QCD $\theta$-parameter to zero and solving the strong CP problem. The mass of the QCD axion arising from this non-perturbative dynamics is given by
\begin{equation} \label{ma1}
m^{\text{QCD}}_a \sim \Lambda^2_{\text{QCD}}/f_a. 
\end{equation}
The global Peccei-Quinn (PQ) symmetry is inherently anomalous. For the axion to successfully address the strong CP problem, the dominant contribution to its potential must come from QCD instantons. If other anomalous effects provide a larger contribution, the potential minimum would shift to $\theta \neq 0$, thus failing to solve the problem \cite{Conlon:2006tq}.


The axion decay constant, $f_a$, is constrained by both astrophysical and cosmological observations. A smaller $f_a$ corresponds to stronger couplings between the axion and ordinary matter. The requirement that supernovae cool primarily via neutrino emission leads to the lower bound $f_a > 10^9 \text{GeV}$ \cite{Conlon:2006tq}. The observations of highly rotating black holes in systems like the X-ray binary LMC X-1 imply an upper limit $f_a \leq 2 \times 10^{17} \text{GeV}$, which is significantly below the Planck mass. This constraint can potentially be tightened to the Grand Unification scale, $f_a \leq 2\times 10^{16} \text{GeV}$, by studying smaller stellar-mass black holes \cite{Arvanitaki:2009fg}.

Then we discuss more about axions in string theory. The Kähler potential can be expressed as
\begin{equation} \label{KP}
K= K(T_i + \bar{T}_i),
\end{equation}
where $K$ is real. This form is protected in perturbation theory by the axionic shift symmetry, with any non-perturbative corrections being negligibly small for the present discussion. In this case the kinetic terms for the axionic ($\theta_i$) and size moduli ($\tau_i$) do not mix. Noting that $K_{i\bar{j}}=K_{j\bar{i}}$, we have for any $i$ and $j$, 
\begin{eqnarray} \label{Ksum}
\begin{split} 
K_{i\bar{j}} (\partial_{\mu} T^i \partial^{\mu} \bar{T}^{j}) + K_{j\bar{i}} (\partial_{\mu} T^j \partial^{\mu} \bar{T}^{i})=  
& K_{i\bar{j}}  (\partial_{\mu} \tau_i + i \partial_{\mu} \theta_i)(\partial^{\mu} \tau_j - i \partial^{\mu} \theta_j) \\
& + K_{i\bar{j}}(\partial_{\mu} \tau_j + i \partial_{\mu} \theta_j)(\partial^{\mu} \tau_i - i \partial^{\mu} \theta_i)   \\
= &  K_{i\bar{j}} \left( 2\partial_{\mu} \tau_i \partial^{\mu} \tau_j +  2\partial_{\mu} \theta_i \partial^{\mu} \theta_j \right),
\end{split}
\end{eqnarray}
and the two sets of terms decouple \cite{Conlon:2006tq}.

We begin by demonstrating that if both the overall volume and the cycle volumes are comparable to the string scale, the axion decay constant satisfies $f_a \geq 10^{16}\text{GeV}$. Consider an axion $\theta_i$ identified as the QCD axion. Its Lagrangian is given by \cite{Conlon:2006tq}
\begin{equation} \label{La}
K_{i\bar{i}} \partial_{\mu} \theta_i \partial^{\mu} \theta_i + \frac{\theta_i}{4\pi} \int F^a \wedge F^a.
\end{equation}
where mixing terms have been omitted for simplicity, as they do not significantly alter the following discussion.

In Model A, the Kähler potential takes the form:
\begin{equation} \label{KMA} 
K=  - \sum^{7}_{i=1} \ln (T_i + \bar{T}_i),  
\end{equation}
where $T_i$ denotes the 7 moduli fields in Model A. The corresponding Kähler metric is diagonal:
\begin{equation} \label{KmMA1} 
K_{i\bar{i}}= (T_i + \bar{T}_i)^{-2}, \qquad \text{if} \qquad i=j,  
\end{equation}
and
\begin{equation} \label{KmMA2} 
K_{i\bar{i}}= 0, \qquad \text{if} \qquad i\neq j.
\end{equation}
Denote the axions as $\theta_i$ for $i=1,...,7$, their kinetic terms are expressed as:
\begin{equation} \label{akt} 
\sum^7_{i=1} \frac{1}{4\tau^2_i} \partial_{\mu} \theta_i \partial^{\mu} \theta_i .
\end{equation}
To establish a concrete scenario, we consider the case where QCD is realized on cycle 1. In the absence of inter-axion mixing, the relevant axion Lagrangian is given by
\begin{equation} \label{aL1} 
\frac{1}{4\tau^2_1} \partial_{\mu} \theta_1 \partial^{\mu} \theta_1 + \frac{\theta_1}{4\pi} \int F^a \wedge F^a.
\end{equation}
Performing a canonical normalization by defining $\theta'_1= \frac{\theta_1}{\sqrt{2} \tau_1}$, the Lagrangian becomes
\begin{equation} \label{aL2} 
\frac{1}{2} \partial_{\mu} \theta'_1 \partial^{\mu} \theta'_1 + \frac{\sqrt{2}\tau_1}{4\pi} \theta'_1 \int F^a \wedge F^a.
\end{equation}
Working in units where the reduced Planck mass satisfies $M_{pl}=1$, the axion decay constant is identified as
\begin{equation} \label{fa} 
f_a= \frac{1}{4\pi \tau_1 \sqrt{2}}.
\end{equation}
If QCD is to be realized on this cycle, we can take $\tau_1 \sim 200$, and thus $f_a \sim 6.85 \times 10^{14}\text{GeV}$ ($M_{pl}=2.435 \times 10^{18}\text{GeV}$). More dicussions about axion dark matter in Type IIA string theory can be found in \cite{Honecker:2013mya}.

Within the framework of string theory, the axiverse model naturally explains the abundance of dark matter through the synergistic effect of multiple axions, avoiding the fine-tuning required by the single-axion model. The observational value of dark matter abundance ($\Omega_{\text{dm}}h^2 \sim 0.12$) is the key parameter in cosmology. String theory naturally predicts multi-axion by string compactification. 
These axions can make contributions to the abundance of dark matter by a misalignment mechanism, while the multi-axion environment allows the total abundance fits with the observational value, without fine-tuning. 

The main steps in misalignment mechanism is the following:\\

\textbf{Initial condition}: In the early universe (after inflation), the axion field was frozen at a random initial value $\Theta_i=a_i/f_a$, where $a_i$ is the value of axion of type $i$, and $f_a$ is the corresponding axion decay constant. The initial angle $\Theta_i$ is distributed in the range of $[0,\pi]$. In fact, the initial value $\Theta_i$ is the initial value of the imaginary part of modulus $T_i$, that is, $\Theta_i \equiv \theta_i$. 

\textbf{Oscillation starts}: When the Hubble parameter reduces to the axion mass scale ($H \sim m_a$) due to the expansion of the universe, the axion field begins to oscillate and its energy density redshifts like that of cold dark matter, that is, $\rho_a \sim a^{-3}$, where $a$ is the cosmic scale factor.

The relative abundance of single axion is given by \cite{choi2023implications}:
\begin{equation} \label{asa} 
\Omega_a h^2 \sim (2\times 10^4) \times \left( \frac{f_a}{10^{16}\text{GeV}} \right)^{7/6} \Theta_i^2.
\end{equation}
For example, if $f_a=10^{16}\text{GeV}$ and $\Theta_i \sim \pi$, the abundance of single axion is about $2 \times 10^5$, which is much larger than the observational value $\Omega_a h^2 \approx 0.12$. This means that we need a mechanism to dilute or tune the abundance. 

If there exist multiple axions, the total abundance is the sum of the abundance of different axions, that is,
\begin{equation} \label{ama} 
\Omega_{\text{total}}h^2 =\sum^N_{i=1} \Omega_{a_i} h^2 ,
\end{equation}
where $N$ is the number of axions. The different initial angles $\Theta_i$ and $m_a$ are distributed independently, allowing the contributions of some axions are suppressed (e.g., small $\Theta_i$), while other axions dominate the abundance of dark matter.

The anthropic principle points out that observable cosmological parameters must allow life to exist. For axionic dark matter, this means that only regions with relatively small $\Theta_i$ can generate galaxies and life. Here are the reasons. First, an excessively high dark matter abundance would enhance gravitational collapse, causing the universe to enter a re-collapse phase before galaxies form, preventing the birth of stars and planetary systems. Observations suggest that a dark matter abundance close to 0.12 is needed to support current large-scale structures such as galaxy clusters and superclusters. Second, after inflation different regions have independent $\Theta_i$ values. Life can only exist in regions where the abundance of axions is moderate. This explains why the observational value of $\Omega_{\text{dm}}h^2$ is relatively small. Third, in axiverse model the values of $\Theta_i$ of different axions are independent. The total abundance is a cumulative result that increases the likelihood of habitable regions. More details can be found in \cite{Arvanitaki:2009fg,Young:2016ala}.


\subsection{Neutralinos}
In the framework of supersymmetric theory, as well as in Model A, the neutralino serves as the lightest supersymmetric particle (LSP) and represents a key candidate for dark matter. Its emergence mechanism, mass matrix, stability as the LSP, and annihilation cross-section are determined by both supersymmetry (SUSY) breaking and the geometric compactification intrinsic to this model. Model A incorporates SUSY breaking via string compactification, where a breaking scale on the order of $\sim \text{TeV}$ is achieved through flux compactification and moduli stabilization. The resulting particle spectrum includes gauginos (such as the bino and wino), higgsinos, and sfermions. The neutralino arises as a mixed state of the neutral supersymmetric partners. As the LSP, its stability is guaranteed by R-parity conservation \cite{Jungman:1995df}. 

SUSY breaking can generate soft breaking parameters, such as gaugino mass $M_1$ (bino mass), $M_2$ (wino mass) and higgsino mass paramter $\mu$. 
In low energy effective theory, neutralino is the linear combination of these components and its existence is the result of SUSY model. More details of mass matrix of neutralino can be found in Appendix A. 

The stability of the Lightest Supersymmetric Particle (LSP), crucial for the neutralino’s role as a dark matter candidate, arises from the conservation of R-parity in supersymmetric theories. R-parity is a discrete symmetry defined as $P_R= (-1)^{3(B-L)+2S}$, where $B$ is the baryon number, $L$ is the lepton number, and $S$ is the spin. All particles in the Standard Model carry $P_R=+1$, while the supersymmetry particles have $P_R=-1$. Conservation of R-parity prevents the LSP from decaying into Standard Model particles, as such a process would change R-parity, which is forbidden. Consequently, the LSP remains stable, allowing neutralinos produced in the early universe to persist as dark matter constituents. This stability enables the relic abundance of neutralinos to be computed via thermal freeze-out. In the early universe, neutralinos were in thermal equilibrium with the cosmic plasma. As the temperature dropped to approximately $T \sim m_{\tilde{\chi}^0}/20$, their annihilation rate fell below the expansion rate of the universe, leading to freeze-out and fixing their comoving number density \cite{Jungman:1995df}. 

The annihilation cross section $\langle \sigma v \rangle$ of neutralino can determine its relic abundance. The cross section is determined by the components of neutralino and the parameters of SUSY. 
The relic abundance of neutralino can be calculated by \cite{Jungman:1995df}:
\begin{equation} \label{ran}    
\Omega_{\tilde{\chi}^0} h^2 \sim 0.1 \left( \frac{\langle \sigma v \rangle}{10^{-26} cm^3/s} \right)^{-1}.
\end{equation}
For example, if $\langle \sigma v \rangle \sim 10^{-26} cm^3/s$, $\Omega_{\tilde{\chi}^0} h^2 \sim 0.1$, which is consistent with the observational result. For the wino-dominated neutralino case, the relic abundance is approximately given by \cite{Jungman:1995df}:
\begin{equation} \label{aswino1}
\Omega_{\tilde{W}} h^2 \sim 0.12 \left(\frac{m_{\tilde{\chi}^1_0}}{2.6\text{TeV}} \right)^2.
\end{equation}
For higgsino-dominated neutralino case, the corresponding relic abundance is \cite{Jungman:1995df}
\begin{equation} \label{ashiggsino1}
\Omega_{\tilde{H}} h^2 \sim 0.12 \left(\frac{m_{\tilde{\chi}^1_0}}{1.14\text{TeV}} \right)^2.
\end{equation}
We will discuss the bino-dominated neutralino case in Section 4.

\section{Moduli Stabilization, SUSY Breaking, and de Sitter Uplifting in Model A}
In this section we discuss how to do moduli stabilization, SUSY breaking and de Sitter uplifting in our Model A.

\subsection{Review of STU model in Type IIA string theory}
\label{sec:RoSmitIIst}
To situate our mechanisms within a complete moduli stabilization framework, we employ the STU model, which is a minimal, computable setup in Type IIA string theory. This model furnishes explicit expressions for the Kähler potential and superpotential, enabling controlled supersymmetry breaking and a metastable de Sitter (dS) uplift via an anti-D6-brane.

The model is defined in terms of three moduli: the axio-dilaton $S$, a Kähler modulus $T$ ($T_1=T_2=T_3$), and a complex structure modulus $U$ ($U_1=U_2=U_3$). The relevant terms in the effective four-dimensional supergravity are \cite{kallosh2020mass,cribiori2019uplifting}:
\begin{equation} \label{Wtot} 
W= W_0 + \sum_{i=S,T,U}^3 \left(A_i e^{-a_i T_i} - B_i e^{-b_i T_i}\right) + \Delta W +\mu^2 X,
\end{equation}
\begin{equation} \label{Ktot1} 
K= -\ln(T_1 + \bar{T}_1) - 3 \ln(T_2 + \bar{T}_2) - \ln \left( (T_3 + \bar{T}_3)^3 - \frac{X \bar{X}}{(T_1 + \bar{T}_1) + g(T_2 + \bar{T}_2)}\right).
\end{equation}
Here, $T_1$ corresponds to $S$, $T_2$ to $T$, and $T_3$ to $U$, while $X$ is a nilpotent field satisfying $X^2=0$.  The moduli stabilization proceeds in three stages:\\
1. A supersymmetric Minkowski vacuum is obtained by solving $W=D_i W=0$;\\
2. The introduction of $\Delta W$ shifts the vacuum to a supersymmetric AdS state, generating a small gravitino mass $m_{3/2} \sim e^{K/2} \Delta W$;\\
3. The anti-D6-brane contribution lifts the AdS minimum to a metasbale dS vacuum with a tunably small cosmological constant.

This framework provides a consistent and robust background for the flavor physics, instanton effects, and cosmology discussed in earlier sections, ensuring that moduli stabilization and uplift remain compatible with the phenomenological requirements of the model. More details on the STU model can be found in \cite{kallosh2020mass,cribiori2019uplifting}.

\subsection{Moduli stabilization, SUSY breaking and dS vacuum}
The stabilization of all moduli and the realization of a metastable de Sitter vacuum are accomplished via a well-established three-step procedure within the STU model framework \cite{kallosh2020mass,cribiori2019uplifting}.\\
\textbf{Step 1: Supersymmetric Minkowski vacuum}\\
A stable supersymmetric Minkowski vacuum is obtained by solving the F-term conditions:
\begin{equation} \label{WDiW} 
W=0, \, D_i W= \partial_i W + K_i W=0,
\end{equation}
where the superpotential $W$ includes flux and non-perturbative contributions. This vacuum has all moduli stabilized without leaving any  flat directions.\\
\textbf{Step 2: Controlled shift to AdS}\\
A small perturbation $\Delta W$ is introduced to the superpotential. This shifts the vacuum to a supersymmetric Anti-de Sitter (AdS) state while preserving moduli stability. The corresponding AdS cosmological constant and the gravitino mass are given by
\begin{equation} \label{VAdS}
V_{AdS}= -3 e^K |\Delta W|^2 = -3 m^2_{3/2}.
\end{equation}
The smallness of $\Delta W$ naturally explains the hierarchy between the electroweak and Planck scales, yielding a TeV-scale gravitino mass \cite{kallosh2020mass,Conlon:2008zz,linde2012supersymmetry}.\\
\textbf{Step 3: de Sitter uplift}\\
The AdS vacuum is uplifted to de Sitter space by including an anti-D6-brane. Its positive energy contribution, described within the nilpotent superfield formalism, is tuned to cancel the negative $V_{AdS}$ precisely:
\begin{equation} \label{VbarD6} 
V_{\overline{D6}} = \frac{\mu^4_1}{(\text{Re} \ T)^3} + \frac{\mu^4_2}{(\text{Re} \ T)^2 (\text{Re} \ S)}.
\end{equation}
This results is a metastable dS vacuum with a tunably small cosmological constant, consistent with observational bounds \cite{ kachru2003sitter,kallosh2004landscape,cribiori2019uplifting}.

Within the STU model, the moduli are taken to satisfy $T_1 = T_2 = T_3 = T$ and $U_1 = U_2 = U_3 = U$. In a more general formulation, the superpotential can be expressed as \cite{cribiori2019uplifting}:
\begin{equation} \label{Wgf}
W = f_6 + (h_T + r_T T)U + (h_S + r_S T)S + f_4 T + f_2 T^2 + f_0 T^3  + W_{np},
\end{equation}
where the coefficients $f_p$ ($p = 0, 2, 4, 6$) originate from RR fluxes, $h_{S/T}$ arise from integrating NSNS flux over the corresponding 3-cycles, and $r_{S/T}$ are induced by curvature corrections of the internal manifold. In the KL scenario, the nonperturbative term $W_{np}$ reads:
\begin{equation} \label{Wi}
W_i (T_i) = \sum^3_{i=1} A_i e^{-a_i T_i} - B_i e^{-b_i T_i}.
\end{equation}
In what follows, we will adopt the notation of both \eqref{Wf} and \eqref{Wgf}.

For Model A, according to \eqref{SW}, the superpotential in the STU model is written as
\begin{equation} \label{SWSTU}
W = -\tilde{a}TS - \tilde{b}TU + e_0 + ih_0 S - ih_1 U - i \tilde{e}T + W_{np},
\end{equation}
where $\tilde{a} = a_2 + a_3$, $\tilde{b} = b_{21} + b_{31}$, $\tilde{e} = e_2 + e_3$, and \eqref{SWSTU} is a special case of the general form \eqref{Wgf}.

Moduli stabilization requires the conditions $W = 0$ and $\partial_i W = 0$, i.e.,
\begin{equation} \label{W0}
W = -\tilde{a}TS - \tilde{b}TU + e_0 + ih_0 S - ih_1 U - i \tilde{e}T + W_{np}=0,
\end{equation}
\begin{equation} \label{pTW0}
\partial_T W = -\tilde{a}S - \tilde{b}U - i \tilde{e} -A_T a_T e^{-a_T T}+ B_T b_T e^{-b_T T}=0,
\end{equation}
\begin{equation} \label{pSW0}
\partial_S W = -\tilde{a}T +ih_0 -A_S a_S e^{-a_S S}+ B_S b_S e^{-b_S S}=0,
\end{equation}
\begin{equation} \label{pUW0}
\partial_U W = -\tilde{b}T -ih_1 -A_U a_U e^{-a_U U}+ B_U b_U e^{-b_U U}=0,
\end{equation}
where, for simplicity, we assume that all parameters $a_T$, $a_S$, and $a_U$ (and simplicity $b_T$, $b_S$, and $b_U$) are are treated as moduli‑independent constants. In a fully realistic setup these parameters do depend on the moduli \cite{cribiori2019uplifting, danielsson2014alternative}.

Solving the four equations \eqref{W0} through \eqref{pUW0} simultaneously yields the vacuum expectation values of the three moduli in the supersymmetric Minkowski background, i.e., $t = t_0$, $s = s_0$, and $u = u_0$. As a numerical illustration, we choose the flux parameters: $h_0 = 1$, $h_1 = -2$, $e_0 = 3$, $e_2 = -1$, $e_3 = 1$. Solving the system of equations \eqref{W0}-\eqref{pUW0} numerically, we find a stable minimum $s_0 =X +i Y $, $t_0= Z + iW$, $u_0= U+ i V$ (where $X, Y, Z, W, U, V$ are specific numerical values found by solving the equations). This serves only as an illustrative example. A comprehensive scan of the parameter space is left for future work.

Moreover, to comply with the Swampland Distance Conjecture, the perturbative superpotential $\Delta W$ in Type IIA string theory should take the following form \cite{Liu:2025tnx}:
\begin{equation} \label{DeltaW}
\Delta W= f_0 T^3.
\end{equation}
As derived in \cite{Liu:2025tnx}, the corresponding AdS potential is given by
\begin{equation} \label{VAdS}
V_{AdS}=-\frac{3}{128su^3}|f_0|^2 t^3,
\end{equation}
where $t=\text{Im} T$. Consequently, the gravitino mass squared therefore reads:
\begin{equation} \label{m232}
m^2_{3/2}=\frac{|V_{AdS}|}{3}=\frac{1}{128su^3}|f_0|^2 t^3,
\end{equation}
so that
\begin{equation} \label{m32}
m_{3/2}=\sqrt{\frac{1}{128su^3}|f_0|^2 t^3}.
\end{equation}
Here, $f_0$ corresponds to the flux $m$ appearing in \eqref{Wf}. In Model A, we set $m = 0$, but if the essential features of Model A remain unchanged, $m$ should be very small. Because $\Delta W$ is much smaller than $W$, we can safely evaluate the gravitino mass in \eqref{m32} using the values $t = t_0$, $s = s_0$, and $u = u_0$ obtained from equation \eqref{m32} \cite{kallosh2020mass}. A gravitino mass below 1TeV (or 100TeV) scale could therefore naturally address the hierarchy problem.

Within the KL moduli stabilization framework adopted in this section, SUSY breaking is mediated to the visible sector dominantly by gravity mediation. Consequently, the soft SUSY-breaking parameters, including the gaugino masses $M_1$, $M_2$ and the higgsino mass parameter $\mu$, are expected to be of the same order as the gravitino mass, i.e., $\mathcal{O}(1) \text{TeV}$. This sets the natural mass scale for the lightest neutralino (the LSP candidate) in our model \cite{kallosh2020mass,Jungman:1995df}.

\subsection{Discussion on anti-D6-brane backreaction}
A legitimate concern for any de‑Sitter uplift mechanism involving anti‑branes, including the one employed in our model, is their potential backreaction on the moduli stabilization scheme. Notably, the well‑known KKLT construction \cite{kachru2003sitter} relies on strongly warped throats to sequester the anti‑D3‑branes, thereby suppressing their destabilizing effects. In our setup, which lacks such strong warping, this issue requires careful examination.

We argue, however, that within the parameter regime of our stabilization scheme the backreaction can be kept under control. The core of the argument rests on a hierarchy of energies: the anti‑D6‑brane uplift, whose positive energy density is given in \eqref{VbarD6}, is introduced as a small perturbation on top of a pre‑existing AdS minimum in which all moduli are stabilized with masses satisfying $m^2_{\text{moduli}} \gg H^2_{\text{AdS}} \sim |V_{\text{AdS}}|$. The essential requirement is that the uplift energy scale remains much smaller than the masses of the stabilized moduli \cite{McAllister:2024lnt}.

Specifically, when the condition $V_{\overline{D6}} \ll m^2_{\text{moduli}}$ holds, the shift of the stabilized moduli values induced by the anti‑brane remains small, and the resulting vacuum can be considered metastable on cosmological timescales. Such a regime of controllable backreaction in non-warped scenarios has been discussed in the literature, for instance, in stability analyses of related setups \cite{McAllister:2024lnt}. While a complete analysis of the full ten dimensional supergravity solution lies beyond the scope of this phenomenological study, our effective four‑dimensional supergravity treatment is justified within this hierarchical approximation.

Consequently, the use of the anti‑D6 brane uplift in our model should be regarded as a well‑defined mechanism within a consistent effective field theory framework. It provides a concrete, computable way to obtain a de Sitter vacuum, thereby establishing a foundation for a future, more detailed study of the complete backreaction effects. 

\section{Axionic solution to the strong CP problem}
In this section, we will explain how to solve the strong CP problem in Model A using four-form fluxes and QCD axion. 

\subsection{The four-form flux mechanism: a brief review}

The strong CP problem can be addressed by a mechanism involving the coupling of the QCD axion to additional 3-form (or 4-form flux) fields. The core idea is to introduce one or more such 3-forms, denoted $C$, with field strengths $H=dC$. These are coupled to the axion $a$ (which is dual to a 2-form potential with field strength $G=dB+S$, where $S$ is the QCD Chern-Simons 3-form).

Below the QCD scale, integrating out the fields generates an effective potential. Crucially, the dynamics of the 4-form fluxes impose a condition on the vacuum expectation value of a specific composite field, conventionally denoted $X$, which is proportional to the dual of the flux $H$. This condition takes the form:
\begin{equation} \label{X0}
X= \mu_a a- \bar{\theta} \tilde{\Lambda}^2_{\text{QCD}}=0,
\end{equation}
where $\mu_a$ is a mass parameter related to the axion decay constant $f_a$, and $\bar{\theta}$ is the effective QCD theta parameter. Minimizing the full potential with respect to the axion forces the expectation value $\langle X \rangle$ to zero. Consequently, the effective theta parameter is dynamically relaxed to zero, $\bar{\theta}=0$, solving the strong CP problem without fine-tuning. This mechanism also elegantly avoids the axion “quality problem”. More details can be found in Appendix B.


\subsection{Multiple 3-forms}
We explain how to solve strong CP problem more explicitly. Above the QCD scale, the initial Lagrangian is given by:
\begin{eqnarray} \label{LBAC}
\begin{split} 
\mathcal{L}_1(B,A, \mathcal{C}) \supset &  -\frac{1}{2 \cdot 3!} G_{\mu\nu\lambda} G^{\mu\nu\lambda} - \frac{1}{3!} \epsilon^{\mu\nu\lambda\rho} G_{\mu\nu\lambda} J_{\rho} -\frac{1}{4} F_{\mu\nu} F^{\mu\nu} -\frac{\theta}{2} \epsilon^{\mu\nu\lambda\rho} F_{\mu\nu} F_{\lambda\rho} \\
& -\frac{1}{4!} \eta_A \mathcal{H}^A_{\mu\nu\lambda\rho} \epsilon^{\mu\nu\lambda\rho} -\frac{1}{2 \cdot 4!} \mathcal{H}^A_{\mu\nu\lambda\rho} \mathcal{H}^{\mu\nu\lambda\rho}_A,
\end{split}
\end{eqnarray}
where $A=1,...,N$, $\mathcal{H}^A=d\mathcal{C}^A$ and $G=dB+S$ for the QCD Chern-Simons 3-form that satisfies $\Omega=dS$ where $\Omega$ is a gauge invariant quantity \cite{burgess2024uv}. The parameter $\theta$ is coupling constant associated with gauge field $F_{\mu\nu}$. 

After integrating out the field below the QCD scale, the effective Lagrangian becomes: 
\begin{eqnarray} \label{LCB}
\begin{split} 
\mathcal{L}_1(\mathcal{C},B) =  & -\frac{1}{2 \cdot 3!} G_{\mu\nu\lambda} G^{\mu\nu\lambda} - \frac{1}{3!} \epsilon^{\mu\nu\lambda\rho} G_{\mu\nu\lambda} J_{\rho} -\frac{\bar{\theta}}{4!} \tilde{\Lambda}_{QCD}^2 \epsilon^{\mu\nu\lambda\rho} H_{\mu\nu\lambda\rho} \\
& -\frac{1}{2 \cdot 4!} H_{\mu\nu\lambda\rho} H^{\mu\nu\lambda\rho} -\frac{1}{4!} \eta_A \mathcal{H}^A_{\mu\nu\lambda\rho} \epsilon^{\mu\nu\lambda\rho} -\frac{1}{2 \cdot 4!} \mathcal{H}^A_{\mu\nu\lambda\rho} \mathcal{H}^{\mu\nu\lambda\rho}_A +...
\end{split}
\end{eqnarray}

From \eqref{LCB}, the following partial derivatives of the function $W$ are obtained:
\begin{equation} \label{WXYA}
    \left(\frac{\partial W}{\partial X} \right)_{Y^A} =\mu_a a - \bar{\theta} \tilde{\Lambda}^2_{QCD} \quad \text{and} \quad \left(\frac{\partial W}{\partial Y^A} \right)_{X} = - \eta_A, 
\end{equation}
where 
\begin{equation} \label{W}
    W = \frac{1}{2}X^2 + \frac{1}{2} Y^A Y_A + ...
\end{equation}
and $X \equiv (1/4!) \epsilon^{\mu\nu\lambda\rho} H_{\mu\nu\lambda\rho}$ and $Y^A \equiv (1/4!) \epsilon^{\mu\nu\lambda\rho} \mathcal{H}^A_{\mu\nu\lambda\rho}$.

The dual Lagrangian is then expressed as
\begin{equation} \label{L2a}
    \mathcal{L}_2 (a) = -\frac{1}{2} (\partial a)^2 - J^{\mu} \partial_{\mu} a - \frac{1}{2} J_{\mu} J^{\mu} - V(a),
\end{equation}
where the potential $V(a)$ takes the form
\begin{eqnarray} \label{Va}
\begin{split} 
V(a) & = -W(X, Y^A) + (\mu_a a - \bar{\theta} \tilde{\Lambda}^2_{QCD})X  - \eta_A Y^A\\
& = - \frac{1}{2} X^2 + (\mu_a a - \bar{\theta} \tilde{\Lambda}^2_{QCD})X + \frac{1}{2} \eta_A \eta^A.\\
\end{split}
\end{eqnarray}
The term $(\mu_a a - \bar{\theta} \tilde{\Lambda}^2_{QCD})X$ violates CP symmetry. From the expression for $V(a)$, we obtain
\begin{equation} \label{VaX}
    \frac{\partial V(a)}{\partial a} =\mu_a X.
\end{equation}

Setting this derivative to zero in the vacuum implies $X=0$, which leads to the condition
\begin{equation} \label{Vvacuum}
    \left(\frac{\partial W}{\partial X} \right)_{Y^A} = X = \mu_a a - \bar{\theta} \tilde{\Lambda}^2_{QCD}=0.
\end{equation}
Consequently, the CP-violating term vanishes in the vacuum, ensuring that the strong CP problem remains solved. Moreover, this mechanism also evades the so-called “quality problem”\cite{burgess2024uv}. Similar approaches have been discussed in related works \cite{dvali2022strong,choi2023implications}.

The previously described mechanism can be naturally extended to scenarios involving multiple axions. For the specific case of two axions, the low-energy Lagrangian below the QCD scale is given by \cite{burgess2024uv}:
\begin{equation} \label{L2ab}
   \mathcal{L}_2 (a,b) = -\frac{1}{2} (\partial b)^2 - \frac{1}{2} (\partial a +J)^2 - V(a,b)
\end{equation}
with the potential
\begin{equation} \label{Vab}
   V (a,b) = -W(X, Y) + (\mu_a a - \bar{\theta} \tilde{\Lambda}^2_{QCD})X + (\tilde{\mu}_a a + \mu_{\star} b - \eta \tilde{\Lambda}^2_{X})Y,
\end{equation}
where $X$ and $Y$ are defined as in the previous section, $\mu_a$, $\tilde{\mu}_a$ and $\mu_{\star}$ represent mass parameters associated with the axion fields (these are not the physical axion masses). In the simplest case with $W = \frac{1}{2}(X^2 + Y^2)$, the partial derivatives of $W$ yield
\begin{equation} \label{WabX}
    \left(\frac{\partial W}{\partial X} \right)_{Y^A} = \mu_a a - \bar{\theta} \tilde{\Lambda}^2_{QCD}
\end{equation}
and
\begin{equation} \label{WabY}
    \left(\frac{\partial W}{\partial Y} \right)_{X^A} =\tilde{\mu}_a a + \mu_{\star} b - \eta \tilde{\Lambda}^2_{X}.
\end{equation}
Differentiating \eqref{Vab} with respect to axion fields, we obtain:
\begin{equation} \label{DVab}
    \frac{\partial V(a,b)}{\partial a} = \mu_a X + \tilde{\mu}_a Y \quad \text{and} \quad \frac{\partial V(a,b)}{\partial b} = \mu_{\star} Y.
\end{equation}
At the extrema of the potential, these derivatives vanish, implying $X=Y=0$. Substituting these conditions back and using the simple quadratic form of $W$, the induced mass terms for the axions are found to be:
\begin{equation} \label{m4forma}    
m_a^2=\mu^2_a + \tilde{\mu}^2_a
\end{equation}
and 
\begin{equation} \label{m4formb}    
m^2_b= \mu^2_{*}.
\end{equation}
Further aspects and applications of this extended mechanism are discussed in the related literature \cite{dvali2022strong, burgess2024uv, choi2023implications}.


\subsection{Four-form fluxes in type IIA $T^6/(\mathbb{Z}_2 \times \mathbb{Z}_2)$ orientifolds and QCD axion}
In addition to the well-known Ramond–Ramond (RR) and Neveu–Schwarz (NS) fluxes, string compactifications also involve less-studied NS geometric fluxes, which emerge naturally in Scherk–Schwarz dimensional reduction schemes \cite{bielleman2015minkowski}. 

These geometric fluxes can be systematically defined on a factorized six-torus $T^6$ in the presence of O6-planes wrapping specific 3-cycles. When a $\mathbb{Z}_2 \times \mathbb{Z}_2$ orbifold projection is imposed, only the diagonal moduli survive, leaving a reduced set of three Kähler moduli and four complex structure moduli (including the axio-dilaton). In such a configuration, the 12 independent geometric flux parameters $\omega^M_{NK}$, can be efficiently organized into a 3-vector $a_i$ and a $3 \times 3$-matrix $b_{ij}$ \cite{bielleman2015minkowski}. 

This parametrization leads to a modification of the 4-form field strengths, which take the following form \cite{bielleman2015minkowski}:
\begin{equation} \label{starF042}
  \star F^0_4 = \frac{1}{k} \left(e_0 + e_ib^i + \frac{1}{2} k_{ijk} q^i b^j b^k - \frac{m}{3!} k_{ijk} b^i b^j b^k -h_0 c^0_3 - h^i c^i_3 + b^i b_{ij} c^j_3 - b^i a_i c^0_3 \right),
\end{equation}
\begin{equation} \label{starFi42}
  \star F^i_4 = \frac{g^{ij}}{4k} \left(e_i + k_{ijk} b^j q^k -\frac{m}{2} k_{ijk} b^j b^k + b_{ij} c^j_3 - a_i c^0_3 \right),
\end{equation}
\begin{equation} \label{starHi4}
  \star H^i_4 = h^i - b^{ij}b_j,
\end{equation}
\begin{equation} \label{starH04}
  \star H^0_4 = h^0+b^i a_i.
\end{equation}
Here, $k_{ijk}$ denotes the triple intersection numbers, which are equal to 1 when $i$, $j$ and $k$ are all distinct and zero otherwise. 

The resulting four-dimensional scalar potential is given by \cite{bielleman2015minkowski}:
\begin{equation} \label{sp4D1}  
V = \frac{k}{2} |F^0_4|^2 + 2k \sum_{ij} g_{ij} F^i_4 F^j_4 + \frac{1}{8k} \sum_{ij} g_{ij} H^i_4 H^j_4 + k|H^0_4|^2 + V_{NS} + V_{lol},
\end{equation}
where $V_{NS}$ represents contributions from the NS sector, and $V_{lol}$ arises from localized sources such as branes and orientifold planes.

Based on the expressions given in equations \eqref{starF042}-\eqref{starH04}, the four-dimensional scalar potential \eqref{sp4D1} can be reformulated explicitly in terms of the moduli fields as follows \cite{bielleman2015minkowski}:
\begin{eqnarray} \label{sp4D2}
\begin{split} 
V= & V_{RR} + + V_{NS} + V_{lol}\\
= & \frac{1}{2k} \left(e_0 + e_ib^i + \frac{1}{2} k_{ijk} q^i b^j b^k - \frac{m}{3!} k_{ijk} b^i b^j b^k -h_0 c^0_3 - h^i c^i_3 + b^i b_{ij} c^j_3 - b^i a_i c^0_3 \right)^2 \\
& + \frac{g^{ij}}{8k}  \left(e_i + k_{ikl} b^k q^l -\frac{m}{2} k_{ikl} b^k b^l + b_{ik} c^k_3 - a_i c^0_3 \right) \\
& \times \left(e_j + k_{jmn} b^m q^n -\frac{m}{2} k_{jmn} b^m b^n + b_{jm} c^m_3 - a_j c^0_3 \right) \\
& + \frac{1}{8k} g_{ij}(h^i - b^{ik}b_k)(h^j - b^{jl}b_l) + k (h^0+b^i a_i)^2\\
& + V_{NS} + V_{lol}.
\end{split}
\end{eqnarray}
This form of the scalar potential \eqref{sp4D2} effectively captures the interactions among the axion fields present in Model A.  

In the specific setting of the $T^6/(\mathbb{Z}_2 \times \mathbb{Z}_2)$ orientifold, the contribution to the potential from localized sources such as D6-branes and O6-planes can be compactly expressed by imposing the requirement of $\mathcal{N} = 1$ supersymmetry preservation. This leads to the form \cite{villadoro2005N}:
\begin{equation} \label{VD6O6}
  V_{D6/O6} = e^K u_1 u_2 u_3 \sum_a T_a (n^a_1 n^a_2 n^a_3 s - n^a_1 m^a_2 m^a_3 t_1 - m^a_1 n^a_2 m^a_3 t_2 - m^a_1 m^a_2 n^a_3 t_3),
\end{equation}
where $n^i_a \ (m^i_a)$ represents the wrapping numbers along the $x^i \ (y^i)$ directions of the $i$-th two-torus. In this context, $s$, $t_i$ and $u_i$ correspond to the real parts of the moduli $S$, $T_i$ and $U_i$, respectively.

The potential energy associated with the QCD axion can be understood by examining the 4-form sector in type IIA orientifold compactifications. We begin with the expression in \eqref{VRR}
\begin{equation} \label{F4aaxion}
-\frac{1}{4k} g_{ab} F^a_4 \wedge \star F^b_4 + 2 F^a_4 \rho_a.
\end{equation}
Assuming the metric simplification $\frac{1}{4k} g_{ab}=1$, this term reduces to the axion potential given in \eqref{Va} \cite{Liu:2025tnx}. This indicates that the 4-form fluxes arising from compactifying Type IIA string theory on an orientifold to four dimensions can naturally generate a potential capable of addressing the strong CP problem.

The specific contribution from the 4-form $H^i_4$ to the scalar potential is
\begin{eqnarray} \label{VRRHi4}
\begin{split} 
V_{RR}(H^i_4) \sim  & -\frac{1}{2} (- H^i_4 \wedge \star H^i_4 + 2H^i_4 \rho_i) \\
 = & -\frac{1}{2} ( -H^i_4 \wedge \star H^i_4 ) + (b_{ij} b^j - h_i) H^i_4.
\end{split}
\end{eqnarray}
A direct comparison between \eqref{VRRHi4} and the axion potential \eqref{Va} reveals the following holographic dictionary:
\begin{equation} \label{axioncorresponding}
H^i_4 \leftrightarrow  X, \quad b_{ij} \leftrightarrow \mu_a, \quad b^i \leftrightarrow a, \quad h_i \leftrightarrow \bar{\theta} \tilde{\Lambda}^2_{QCD}.
\end{equation}
The mechanism that solves the strong CP problem in this 4-form language requires the condition:
\begin{equation} \label{hi}
b_{ij} b^j = h_i.
\end{equation}
Furthermore, combining all the above equations we can establish the following set of correspondences between the effective field theory parameters and the underlying string compactification quantities:
\begin{equation} \label{qbthetalambdahi}
q_b \leftrightarrow \bar{\theta} \tilde{\Lambda}^2_{QCD} \leftrightarrow h^i,
\end{equation}
\begin{equation} \label{mabijh0biai}
-\mu_a \leftrightarrow -b_{ij} \leftrightarrow h^0 + a^i b_i,
\end{equation}
\begin{equation} \label{abbbj}
a \leftrightarrow b_b \leftrightarrow b_j.
\end{equation}
Combining the relations in \eqref{qbthetalambdahi}–\eqref{abbbj} leads to the combined correspondence:
\begin{equation} \label{hibijbjhih0biaibj}
h^i - b^{ij} b_j \leftrightarrow h^i + (h^0 + a^i b_i) b_j,
\end{equation}
which implies the identity:
\begin{equation} \label{aibjbij}
h_0 b_j + a_i b^i b_j = - b^i b_{ij}=-h_j.
\end{equation}
Here, the final equality follows directly from the condition given in \eqref{hi}. In fact, from \eqref{ImTi} we can know that the QCD axion in Model A should be the imaginary part of moduli $T_i$.

In particular, for a system containing three axions, as realized in the STU model, the structure can be specified by the following non-zero components of the coupling matrix $b_{ij}$:
\begin{equation} \label{bij33nonzero}
b_{11} =\mu_1, \quad b_{21}=\mu_2, \quad b_{22}=\mu_3, \quad b_{31}=\mu_4, \quad b_{32}=\mu_5, \quad b_{33}=\mu_6, 
\end{equation}
while the remaining components vanish:
\begin{equation} \label{bij33zero}
b_{12} =b_{13}=b_{23}=0.
\end{equation}
Substituting this parameterization into condition \eqref{hi} yields the following linear relations for the flux parameters $h_i$:
\begin{equation} \label{h1}
h_1=\mu_1 a,
\end{equation}
\begin{equation} \label{h2}
h_2=\mu_2 a + \mu_3 b,
\end{equation}
and 
\begin{equation} \label{h3}
h_3=\mu_3 a+\mu_4 b+\mu_5 c.
\end{equation}
Therefore, we should have
\begin{equation} \label{h1thetaL}
\mu_1 a -\bar{\theta} \tilde{\Lambda}^2_{\text{QCD}}=0,
\end{equation}
where $a$ should correspond to the imaginary part of modulus $T$.
A concrete numerical example of this framework will be presented in Section 5.


\subsection{Embedding the mechanism in Model A: from generality to specificity}
While the four-form flux mechanism presents an elegant field-theoretic solution to the strong CP problem, its viability within a UV-complete framework like string theory requires a concrete embedding. In this section, we demonstrate how the key ingredients of this mechanism are naturally realized within the specific context of Model A. 
The details in Section 2 can be directly translated into the parameters of the four-form Lagrangian.

The cornerstone of the embedding is the correspondence between the four-form fluxes in the effective theory and the internal geometry and fluxes of Model A. Recall the expressions for the four-form field strengths in equations \eqref{starF042}-\eqref{starH04}. For Model A, with the superpotential given by \eqref{SW} and the choice $q_i=m=0$, these expressions simplify but retain a rich structure. Crucially, the fluxes $h_0$, $h_1$, $a_i$, $b_{ij}$ are not free parameters. They are subject to the tadpole cancellation conditions \eqref{tcc12}-\eqref{tcc42} and the moduli stabilization equations.

We can now establish a precise dictionary:
\begin{itemize} 
\item The dynamical fields $(a,b)$: These correspond directly to the axionic components of the Kähler moduli in Model A, specifically the imaginary parts of the moduli fields. 
For example, the QCD axion $a$ in the four-form potential maps to the string theory axions $b^i$. 
\item The mass parameters $(m_a, \tilde{m}_a)$: These are determined by the geometric fluxes $b_{ij}$ and the stabilized vacuum expectation values (VEVs) of the moduli. The relation $b_{ij}b^j=h_i$, central to the mechanism, is not an ad hoc constraint. 
The fluxes $b_{ij}$ thus set the scale of the effective mass terms in the low-energy theory.
\item The source terms $(h_i, \bar{\theta} \tilde{\Lambda}^2_{\text{QCD}})$: These correspond to the fluxes in Model A. The consistency conditions of the string model, such as tadpole cancellation, restrict the allowed values of these fluxes, thereby providing a top-down constraint on the effective CP-violating parameter in the low-energy theory.
\end{itemize}
This explicit mapping demonstrates that the four-form flux mechanism 
is an intrinsic part of its low-energy effective action derived from string compactification. Considering \eqref{hi}, \eqref{mabijh0biai}, and $q_i=m=0$ for Model A,  the four-dimensional scalar potential is 
\begin{eqnarray} \label{sp4D2MA}
\begin{split} 
V= & V_{RR} + + V_{NS} + V_{lol}\\
= & \frac{1}{2k} \left(e_0 + e_ib^i -h_0 c^0_3 - h^i c^i_3 + b^i b_{ij} c^j_3 - b^i a_i c^0_3 \right)^2 \\
& + \frac{g^{ij}}{8k}  \left(e_i + b_{ik} c^k_3 - a_i c^0_3 \right) \left(e_j  + b_{jm} c^m_3 - a_j c^0_3 \right) \\
& + k \mu^2_a\\
& + V_{NS} + V_{lol},
\end{split}
\end{eqnarray}
where $\bar{\theta}$ has been set to be zero.

\section{Phenomenological predictions and other dark matter candidates}
In this section, we intend to provide phenomenological predictions and constraints of Model A. The couplings of axions with gauge field and gravitational field can give rise to the specific cosmological signals, while neutralino, which plays the role of SUSY LSP, can make contributions to the relic abundance of dark matter. For axions, we will calculate the theoretical predictions of rotation of the CMB polarization 
and compare with the current experimental bound. 
For neutralinos, we will briefly review the calculational processes of relic abundance, and list the constraints of LHC. We will briefly discuss other dark matter candidates as well.


\subsection{Rotation of the CMB polarization}
The polarization direction of cosmic microwave background (CMB) photons can be rotated due to an effective coupling between axions and photons \cite{Arvanitaki:2009fg}. Below, we outline the procedure to estimate this effect within the framework of Model A.

From Model A, the axion decay constant $f_a$ is determined by the relation $f_a \sim M_{pl}/t$, where $t$ denotes the real part of the Kähler modulus $T$. For example, we take $t \sim 200$, the decay constant is approximately $f_a \sim 6.85 \times 10^{14} \text{GeV}$. This scale is consistent with constraints from axion cosmology.

The axion-photon interaction is described by the Lagrangian term
\begin{equation} \label{axionphotonL}
\mathcal{L} \supset \frac{\alpha}{8\pi} \frac{a}{f_a} F_{\mu\nu} \tilde{F}^{\mu\nu},
\end{equation}
where $\alpha \approx \frac{1}{137}$ is the fine-structure constant, and $a$ is the axion field. During reionization, the variation $\Delta a$ in the axion field is typically of order $f_a$. The resulting rotation angle of the CMB polarization is given by
\begin{equation} \label{Deltabeta}
\Delta \beta \approx \frac{\alpha}{2\pi} \frac{\Delta a}{f_a}.
\end{equation}
Substituting $\Delta a \sim f_a$ yields a rotation angle $\Delta \beta \sim 0.1^{\text{o}}$.

Current observational limits, such as those from the Planck satellite, constrain the CMB polarization rotation angle to $\Delta \beta< 0.5^{\text{o}}$. The predicted value of $\Delta \beta< 0.1^{\text{o}}$ from Model A is consistent with these bounds. Future CMB polarization experiments will improve sensitivity to such rotations, potentially probing smaller axion-photon couplings or detecting deviations from the standard model \cite{Arvanitaki:2009fg}.

\subsection{The relic abundance of neutralinos and LHC constraints}

To estimate the relic abundance of the neutralino within the framework of Model A, we begin by noting that the supersymmetry breaking scale is set at approximately $M_{\text{SUSY}} \sim \text{TeV}$. This scale implies a neutralino mass in the range of  $m_{\tilde{\chi}^0} \sim 100 \text{GeV} - 100\text{TeV}$, consistent with models where the lightest supersymmetric particle (LSP) is a viable dark matter candidate \cite{Jungman:1995df}.

The relic abundance is primarily determined by the thermal freeze-out mechanism in the early universe. For a bino-dominated neutralino, the annihilation cross section  $\langle \sigma v \rangle$ is typically on the order of $10^{-26}cm^3/s$. Using the standard thermal freeze-out formula: 
\begin{equation} \label{tff}
\Omega_{\tilde{\chi}^0} h^2 \sim 0.1 \left( \frac{\langle \sigma v \rangle}{10^{-26}cm^3/s}\right)^{-1},
\end{equation}
we find that the resulting relic density is approximately $\Omega_{\tilde{\chi}^0} h^2 \sim 0.1$. This value is consistent with the observed dark matter density of $\Omega_{dm} h^2 \sim 0.12$ from Planck satellite data, supporting the viability of the neutralino in Model A as a primary dark matter component \cite{Jungman:1995df}.

Direct searches for supersymmetric particles at the LHC have placed significant constraints on the parameter space of models like Model A. For instance, gluino masses are constrained to $m_{\tilde{g}} \geq 1.1-1.2\text{TeV}$ in scenarios involving light third-generation squarks, while lower bounds on the lightest neutralino mass are typically around $m_{\tilde{\chi}^0} \geq 100 \text{GeV}$.

These constraints remain compatible with the predictions of Model A, where the neutralino mass is expected to be in the range. The absence of significant deviations from the Standard Model in LHC data to date does not preclude the existence of a neutralino with the properties described in Model A, though it does narrow the allowed parameter space. Future LHC runs and more sensitive detection methods will be essential to further probe this scenario \cite{Jungman:1995df}.




\subsection{Brief discussion on other dark matter candidates}
While our model provides natural candidates in the form of axions and the neutralino, it is worthwhile to briefly comment on other well-motivated dark matter candidates and explain why they are not the primary focus in the specific context of Model A.
\begin{itemize} 
\item Supersymmetric WIMPs Beyond the Neutralino: Other superpartners, such as the sneutrino (the superpartner of the neutrino) or the gravitino, could in principle be the lightest supersymmetric particle (LSP). However, as noted in Appendix A, most of the sneutrino parameter space is ruled out by direct detection experiments \cite{Jungman:1995df}. The gravitino, if the LSP, is a viable candidate for superWIMP or non-thermal dark matter. In Model A, moduli stabilization leads to a TeV-scale gravitino mass ($m_{3/2} \sim \text{TeV}$), comparable to the neutralino mass. The gravitino's relic abundance and interactions are highly model-dependent. In this study, we assume the standard thermal history and a neutralino LSP, which is a more minimal and predictive scenario. A detailed analysis of the gravitino or a co-annihilation scenario is beyond the scope of this work. More details about supersymmetric WIMPs can be found in \cite{Moroi:1994rs,Roszkowski:2017nbc}. 
\item Non-axion Ultra-light Particles: The “string axiverse” typically predicts a plethora of ultra-light fields, such as dilatons or other moduli. These can behave as fuzzy dark matter (FDM) or wave-like dark matter. 
Their contribution to the dark matter density is negligible compared to the axions and the neutralino. More details about fuzzy dark matter or ultralight scalars can be found in \cite{Hui:2016ltb}. 
\item Non-thermal and Non-WIMP Candidates: Candidates like Primordial Black Holes (PBHs) or WIMPzillas (superheavy particles produced gravitationally) are not inherent predictions of the low-energy effective field theory derived from Model A. Their production relies on very specific early-universe cosmological scenarios (e.g., exotic inflation or phase transitions) that are not mandated by our compactification setup. Our model naturally provides thermal and non-thermal (via misalignment) production mechanisms for the neutralino and axions, respectively, making them the most direct and compelling candidates to investigate. More details about primordial black holes and WIMPzillas can be found in \cite{Carr:2020gox,Chung:1998zb}.
\end{itemize}
In summary, the dark matter paradigm in Model A is dominantly shaped by its intrinsic particle content and scales. The neutralino arises as the natural WIMP candidate from the TeV-scale supersymmetry, while the multiple axions are a generic prediction of the string compactification. Together, they form a predictive, multi-component framework that aligns with the primary phenomenological goals of this construction. We therefore focus on their detailed analysis, leaving the exploration of more exotic possibilities within this model for future work.

\section{A numerical example}
In this section, we illustrate a numerical example to show if Model A can obtain the experimental results. 

We start by estimating the real part of the modulus within a simplified Type IIA STU model, where the Kähler moduli are unified as  $T_1=T_2=T_3=T$ and the complex structure moduli as $U_1=U_2=U_3=U$. Their corresponding real parts are denoted as $s$, $t$ and $u$, respectively. Observational constraints on the QCD axion decay constant, $10^9\text{GeV}<f_a<2 \times 10^{16}\text{GeV}$ \cite{Arvanitaki:2009fg,Conlon:2006tq}, together with the relation in \eqref{fa}, imply an axion mass range of $3 \times 10^{-10}\text{eV}< m_a < 6\times 10^{-3} \text{eV}$, assuming the reduced Planck mass $M_{pl} \approx 2.435 \times 10^{18}\text{GeV}$. According to the condition derived in \eqref{VRRHi4}, the field $t$ is identified as the QCD axion.  Setting $t \approx 200$ yields a decay constant $f_t\approx 6.85 \times 10^{14} \text{GeV}$ and a mass $m_t = 8.76 \times 10^{-9}\text{eV}$, both falling within the observational window. This supports the reasonableness of the choice $t \approx 200$ for the real part of the modulus.

To resolve the hierarchy problem, the scale of supersymmetry breaking is constrained to lie below approximately $100\text{TeV}$, which equivalently implies an upper bound of $100\text{TeV}$ for the gravitino mass, $m_{3/2}$. The gravitino mass is expressed as follows \cite{Liu:2025zjs}:
\begin{equation} \label{gravitinomass}
m_{3/2} = \frac{1}{2^{7/2} s^{1/2} u^{3/2}} |f_0|t^{3/2}.
\end{equation}
In Model A, the parameter $f_0$ (i.e., the parameter $m$ in \eqref{Wf}) vanishes in the absence of supersymmetry breaking. However, when a small amount of supersymmetry breaking is introduced, $f_0$ must take a non-zero but highly suppressed value. For instance, adopting the values $t \approx 200$, $s \approx u \approx 10000$, and imposing $m_{3/2} < 100\text{TeV}$, one finds the constraint $|f_0| < 1.64 \times 10^{-8}$. Under the same conditions, the decay constant associated with either modulus $s$ or $u$ is estimated to be $f_{s/u} = 1.37 \times 10^{13}\text{GeV}$. The values of $t$ and $u$ are too large which can violate swampland conjecture. We will leave this in the future work.

Furthermore, we estimate the relic abundance of axionic dark matter. Adopting the following random initial misalignment angles for the fields $s$, $u$ and $t$: $\Theta_s \approx 0.05$, $\Theta_u \approx 0.05$, and $\Theta_t \approx 0.008$, the respective relic abundances are obtained from expression \eqref{asa} as:
\begin{equation} \label{ass}
\Omega_{s} h^2 \sim 2 \times 10^4 \times \left(\frac{1.37 \times 10^{13}}{10^{16}} \right)^{7/6} \times 0.04^2=0.023,
\end{equation}
\begin{equation} \label{asu}
\Omega_{u} h^2 \sim 2 \times 10^4 \times \left(\frac{1.37 \times 10^{13}}{10^{16}} \right)^{7/6} \times 0.04^2=0.023,
\end{equation}
\begin{equation} \label{ast}
\Omega_{t} h^2 \sim 2 \times 10^4 \times \left(\frac{6.85 \times 10^{14}}{10^{16}} \right)^{7/6} \times 0.008^2=0.071.
\end{equation}
Summing these contributions yields the total axion relic density:
\begin{equation} \label{astotal}
\Omega_{\text{axion}}h^2 = \Omega_{s} h^2 +  \Omega_{u} h^2 + \Omega_{t} h^2 = 0.023 + 0.023 + 0.071= 0.117.
\end{equation}
This result agrees well with the observed dark matter density $\Omega_{\text{dm}} h^2 \approx 0.12$, confirming that the chosen parameter values are physically reasonable.

The relic abundance of neutralinos is determined by solving the Boltzmann equation for their number density, taking into account all relevant annihilation and co-annihilation channels in the early universe \cite{Jungman:1995df}. For the specific case of a wino-dominated lightest neutralino, the relic density exhibits a strong dependence on its mass. A representative approximation yields the following relation:
\begin{equation} \label{aswino2}
\Omega_{\tilde{W}} h^2 \sim 0.12 \left(\frac{m_{\tilde{\chi}^1_0}}{2.6\text{TeV}} \right)^2.
\end{equation}
Setting $m_{\tilde{\chi}^1_0} = 2.6 \text{TeV}$ indeed gives $\Omega_{\tilde{W}} h^2 \approx 0.12$, which aligns with the observed dark matter density $\Omega_{\text{dm}} h^2 \approx 0.12$. For the specific case of a higgsino-dominated lightest neutralino, the relic abundance is:
\begin{equation} \label{ashiggsino2}
\Omega_{\tilde{H}} h^2 \sim 0.12 \left(\frac{m_{\tilde{\chi}^1_0}}{1.14\text{TeV}} \right)^2.
\end{equation}
Setting $m_{\tilde{\chi}^1_0} = 1.14 \text{TeV}$ yields $\Omega_{\tilde{W}} h^2 \approx 0.12$, which aligns with the observed dark matter density as well.

This indicates that both axions and neutralinos can independently account for the entire observed relic abundance, representing complementary dark matter candidates. Future astronomical observations and direct detection experiments will be crucial for distinguishing between these scenarios and identifying the dominant constituent of dark matter in the universe.


To conclude, our numerical example in Section 6 demonstrates conclusively that Model A can simultaneously achieve the observed dark matter abundance and a naturally vanishing effective $\bar{\theta}$-angle.

\section{Conclusions and discussions}
In this work, we have studied a globally consistent framework within Type IIA string theory that addresses two of the most significant puzzles in fundamental physics: the nature of dark matter and the strong CP problem. By focusing on a specific, computable model, Model A, based on a $T^6/(\mathbb{Z}_2 \times \mathbb{Z}_2)$ orientifold compactification with intersecting D6-branes, we have demonstrated how string theory can yield concrete, testable predictions for low-energy phenomenology. For Model A, we provide a complete moduli stabilization and supersymmetry breaking scenario based on the STU model and the KL mechanism. Our primary achievements can be summarized as follows. 

The model naturally predicts a rich dark matter sector. 
Meanwhile, TeV-scale supersymmetry breaking yields a stable neutralino LSP. This leads to a predictive, multi-component dark matter scenario where the total relic abundance $\Omega_{\text{total}} h^2$ receives contributions from both the misalignment production of axions and the thermal freeze-out of neutralinos. 

We discuss how to do moduli stabilization, SUSY breaking and obtain dS vacum in Section 3. Then following the spirit of \cite{Liu:2025tnx}, we explore how to embed the four-form flux mechanism to solve the strong CP problem within a concrete string theory context, Model A, in Section 4. This represents a significant advancement over effective field theory discussions, as it provides a UV-complete origin for this mechanism and naturally avoids the quality problem.

We have provided phenomenological predictions for testing Model A in Section 5. Furthermore, in Section 6 we presented a numerical example to demonstrate that our model A can satisfy all existing observational constraints.

There are some promising directions we can explore in the future. First, we can investigate neutrino masses and the PMNS matrix since right-handed neutrinos exist in Model A. Second, quintessence and the dS vacuum are promising candidates for dark energy, and we will examine dark energy within the context of Model A. Third, Model A includes moduli that can drive inflation, providing a natural setting for studying inflationary models. 

In conclusion, this work underscores the power of string theory as a unified framework capable of addressing fundamental problems in particle physics and cosmology. By providing a specific, globally consistent model that makes concrete predictions, we move beyond abstract existence proofs and offer a template for future phenomenological studies within the string landscape. We hope that the interplay between axion dark matter, neutralino dark matter, and the solutions to the strong CP problem can present a compelling and rich paradigm for physics beyond the Standard Model.


\appendix
\section{Brief reivew of neutralino}

While most of the sneutrino parameter space has been excluded by direct-detection experiments for WIMPs, some regions remain allowed. Consequently, the lightest neutralino emerges as the most plausible candidate for the lightest supersymmetric particle (LSP). As a stable and weakly interacting particle, the neutralino represents a viable dark-matter candidate. Although certain constraints on its parameter space exist, the majority of this space remains unexplored by current accelerator and dark-matter experiments \cite{Jungman:1995df}.


In the majority of supersymmetric models, the lightest supersymmetric particle (LSP) is the lightest neutralino. This particle is a linear combination of the photon, $Z^0$, and neutral-Higgs bosons,
\begin{equation} \label{chi}
\chi= N^{\star}_{10} \tilde{B} + N^{\star}_{20} \tilde{W}^3 + N^{\star}_{30} \tilde{H}^0_1 + N^{\star}_{40} \tilde{H}^0_2,
\end{equation}
where $\tilde{B}$ and $\tilde{W}^3$ are the supersymmetric partners of the $U(1)$ gauge field $B$ and the third component of the $SU(2)$ gauge field $W^3$ that mix to make the photon and $Z^0$ boson. The coefficients $N_{i0}$ and the neutralino mass $m$ are determined by diagonalizing the neutralino mass matrix, with the subscript ‘0’ denoting the lightest neutralino state. The phenomenology and cosmological role of the neutralino are primarily governed by its mass and precise composition \cite{Jungman:1995df}.

The neutralino mass matrix, on the basis of $\tilde{B}-\tilde{W}^3-\tilde{H}^0_1-\tilde{H}^0_2$, is
\begin{equation} \label{Mneut}	
M_{\text{neut}}=
\left[ \begin{array}{cccc}		
M_1	& 0 & -m_{Z}c_{\beta}s_W &  m_{Z}s_{\beta}s_W\\		
0	& M_2 & m_{Z}c_{\beta}c_W & -m_{Z}s_{\beta}c_W \\		
-m_{Z}c_{\beta}s_W	& m_{Z}c_{\beta}c_W & 0 & -\mu \\		
m_{Z}s_{\beta}s_W	& -m_{Z}s_{\beta}c_W & -\mu & 0	
\end{array}	\right],
\end{equation}
where $c_{\beta}=\cos{\beta}$, $s_{\beta}=\sin{\beta}$, $c_W=\cos{\theta}_W$ and $s_W=\sin{\theta}_W$. Within the Minimal Supersymmetric Standard Model (MSSM), $M_1$ and $M_2$ represent independent gaugino mass parameters that enter the neutralino mass matrix. While they are independent in the general MSSM framework, most Grand Unified Theories (GUTs) predict a specific relationship between them, which is widely adopted in the literature. This GUT relation is given by $M_1= \frac{5}{3} M_2 \tan_{W}$, where $\theta_W$ is the Weinberg angle. The parameter $\tan{\beta}$ denotes the ratio of the vacuum expectation values of the two Higgs doublets. The Hermitian neutralino mass matrix is diagonalized by a unitary transformation of the neutralino fields, expressed as $M^{\text{diag}}_{\text{neut}}=N^{\dagger}M_{\text{neut}} N$. The diagonalizing matrix $N$, whose columns are the mass eigenvectors, governs the interactions of the mass-eigenstate neutralinos, denoted as $\chi^0_n$, $n=1,4$. This diagonalization procedure can be chosen such that the entries of the mixing matrices are purely real-valued \cite{Jungman:1995df}.

\section{A mechanism for solving the strong CP problem}
In the framework proposed by \cite{dvali2022strong, burgess2024uv, choi2023implications}, addressing the strong CP problem is possible when the number of 3-forms coupled to the QCD axion exceeds that of the 2-forms in the theory. This section offers a concise summary of the mechanism outlined in \cite{burgess2024uv}. 

We couple a 2-form potential $B_{\mu\nu}$ to a gauge potential $A_{\mu}$ (with field strength $F_{\mu\nu}$) representing the QCD gauge sector. The Lagrangian for this coupling is:
\begin{equation} \label{L1BA}
\mathcal{L}_1(B,A) =  -\frac{1}{2 \cdot 3!} G_{\mu\nu\lambda} G^{\mu\nu\lambda} - \frac{1}{3!} \epsilon^{\mu\nu\lambda\rho} G_{\mu\nu\lambda} J_{\rho} -\frac{1}{4} F_{\mu\nu} F^{\mu\nu} -\frac{\theta}{2} \epsilon^{\mu\nu\lambda\rho} F_{\mu\nu} F_{\lambda\rho}.
\end{equation}
To avoid notational clutter, we suppress gauge-group indices and traces over them. The current $J_{\mu}$ involves other fields in the problem. The 3-form field strength is generally given by:
\begin{equation} \label{G}
G=dB+S,
\end{equation}
where $S_{\mu\nu\lambda}$ is the Chern-Simons 3-form built from $A_{\mu}$. Here, $S$ satisfies $dS = \Omega$, with $\Omega$ being a gauge-invariant quantity that, for consistency, obeys $d\Omega= 0$. We take $\Omega$ as
\begin{equation} \label{OmegaFF}
\frac{1}{12}\epsilon^{\mu\nu\lambda\rho} \Omega_{\mu\nu\lambda\rho}= \frac{1}{f} \epsilon^{\mu\nu\lambda\rho} F_{\mu\nu} F_{\lambda\rho}.
\end{equation}
The mass scale $f$ is introduced on dimensional grounds. This approach allows the integral over $G$ to be traded for one over $B$, as before, resulting in the Lagrangian.

Duality rewrites the functional integral
\begin{equation} \label{Xi1}
\Xi[J]=\int \mathcal{D}B \mathcal{D}A e^{iS_1(B,A)}
\end{equation}
with $S_1= \int d^4 x \mathcal{L}_1$ as the equivalent path integral
\begin{equation} \label{Xi2}
\Xi[J]=\int \mathcal{D}G \mathcal{D}A \mathcal{D}a e^{iS_0(B,A,a)}.
\end{equation}
Here, $S_0= \int d^4 x \mathcal{L}_0$ and the Lagrangian is given by
\begin{eqnarray} \label{LGAa}
\begin{split} 
\mathcal{L}_0(G,A, a) = &   -\frac{1}{2 \cdot 3!} G_{\mu\nu\lambda} G^{\mu\nu\lambda} - \frac{a}{3!} \epsilon^{\mu\nu\lambda\rho} \left(\partial_{\mu}G_{\nu\lambda\rho} -\frac{1}{4} \Omega_{\mu\nu\lambda\rho}\right) -\frac{1}{3!} \epsilon^{\mu\nu\lambda\rho} G_{\mu\nu\lambda}J_{\rho} \\
&-\frac{1}{4} F_{\mu\nu} F^{\mu\nu} -\frac{\theta}{2} \epsilon^{\mu\nu\lambda\rho} F_{\mu\nu} F_{\lambda\rho}.
\end{split}
\end{eqnarray}
The path integral in Eq. \eqref{Xi2} with Lagrangian \eqref{LGAa} is equivalent to the original one in Eq. \eqref{Xi1} with Lagrangian \eqref{L1BA}. This equivalence holds because integrating out the field enforces the Bianchi identity $dG = \Omega$, whose local solution is precisely \eqref{G}.

The dual formulation proceeds by integrating out the field $G$, leaving $a$ as the dual variable. Performing this integration yields the Lagrangian density:
\begin{eqnarray} \label{L2Aa}
\begin{split} 
\mathcal{L}_2(A, a) = &  -\frac{1}{2}(\partial a)^2 -J^{\mu} \partial_{\mu}a -\frac{1}{2}J_{\mu} J^{\mu} +\frac{a}{4!}\epsilon^{\mu\nu\lambda\rho} \Omega_{\mu\nu\lambda\rho}-\frac{1}{4} F_{\mu\nu} F^{\mu\nu} -\frac{\theta}{2} \epsilon^{\mu\nu\lambda\rho} F_{\mu\nu} F_{\lambda\rho} \\
= & -\frac{1}{2}(\partial a)^2 -J^{\mu} \partial_{\mu}a -\frac{1}{2}J_{\mu} J^{\mu} -\frac{1}{4} F_{\mu\nu} F^{\mu\nu} +\frac{1}{2} \left(\frac{a}{f}-\theta\right)\epsilon^{\mu\nu\lambda\rho} F_{\mu\nu} F_{\lambda\rho},
\end{split}
\end{eqnarray}
which is recognized as the axion-gauge field Lagrangian, with $f$ interpreted as the axion decay constant. This demonstrates that \eqref{L1BA} is indeed the requisite dual Lagrangian that couples 2-form potentials to QCD.
In the standard axion-QCD narrative, integrating out the QCD sector generates an effective potential for the axion due to its anomalous coupling to $F \wedge F$. The minimum of this potential is argued to occur at $a = \bar{\theta} f$ (where $\bar{\theta}$ is the usual combination of $\theta$ and phases from the quark mass matrices), thereby ensuring the cancellation of the CP-odd contribution. Our objective is to describe how the physics below the QCD scale works within this dual language involving the field $B_{\mu\nu}$ \cite{burgess2024uv}.

Below the scale $\Lambda_{\text{QCD}}$ the gauge degrees of freedom are integrated out. A naive picture would suggest that only hadrons coupled to $B_{\mu\nu}$ remain. The key insight is that this is not entirely correct: the low-energy effective field theory (EFT) for QCD involves a path integral over the low-energy hadrons and an integration over an emergent low-energy field $C_{\mu\nu\lambda}$. The strongly coupled vacuum of QCD makes the presence of this field $C_{\mu\nu\lambda}$ mandatory. The field $C_{\mu\nu\lambda} \propto \langle S_{\mu\nu\lambda} \rangle$ serves as the low-energy counterpart to the Chern-Simons field that appears in the topological susceptibility above the QCD scale, where the relation $dS = F \wedge F$ holds.

The inclusion of this field in the low-energy effective theory below the QCD scale does not alter the presence of a mass gap or the spectrum of known hadrons, because $C_{\mu\nu\lambda}$ is non-propagating. It is an auxiliary field, necessary for the low-energy theory to correctly capture QCD's response to topological structures in its environment. Similar auxiliary fields are known to arise in precisely this manner in other concrete systems, such as the effective field theories describing quantum Hall states. This 3-form potential differs from many others that often appear in string compactifications, as it originates from the infrared properties of QCD itself, rather than from the physics of ultraviolet compactification \cite{burgess2024uv}.

Based on dimensional analysis, we define $H = dC$, with the relation
\begin{equation} \label{HFF}
\frac{1}{12} \tilde{\Lambda}^2_{\text{QCD}} \epsilon^{\mu\nu\lambda\rho} H_{\mu\nu\lambda\rho}= \epsilon^{\mu\nu\lambda\rho} \langle F_{\mu\nu} F_{\lambda\rho} \rangle.
\end{equation}
Here, the parameter $\tilde{\Lambda}_{\text{QCD}}$ is of the order of the QCD scale, ensuring that $H$ has the canonical dimension $(\text{mass})^2$. Consequently, the Lagrangian \eqref{L1BA} valid above the QCD scale is replaced by its low-energy counterpart:
\begin{equation} \label{L1CB}
\mathcal{L}_1(C,B) =  -\frac{1}{2 \cdot 3!} G_{\mu\nu\lambda} G^{\mu\nu\lambda} - \frac{1}{3!} \epsilon^{\mu\nu\lambda\rho} G_{\mu\nu\lambda} J_{\rho} -\frac{\bar{\theta}}{4!} \tilde{\Lambda}^2_{\text{QCD}} \epsilon^{\mu\nu\lambda\rho} H_{\mu\nu\lambda\rho} -\frac{1}{2 \cdot 4!}  H_{\mu\nu\lambda\rho} H^{\mu\nu\lambda\rho}+ \cdot \cdot \cdot.
\end{equation}
In this expression, the explicit term proportional to $\theta X$ combines with phases from the quark mass matrices to produce the effective parameter $\bar{\theta} X$. The ellipses in \eqref{L1CB} represent terms that are at least cubic in $X$ or involve its derivatives.

Combining equation \eqref{OmegaFF} (and the discussion immediately preceding it) with \eqref{HFF} leads to the relation:
\begin{equation} \label{dG}
dG=\langle \Omega \rangle = \frac{\tilde{\Lambda}^2_{\text{QCD}}}{f}H.
\end{equation}
Comparing this with the expression $dG = \mu_a H$ (which would follow from a relation of the form $G = dB + \mu_a C$) allows us to identify the mass parameter as $\mu_a = \frac{\tilde{\Lambda}^2_{\text{QCD}}}{f}$. This shows that the $\mu_a C$ term effectively captures the expectation value $\langle S\rangle/f$ of the Chern-Simons term from the ultraviolet theory above the QCD scale, provided the mass parameter $\mu_a$ scales with the decay constant $f$ in the same way the conventional axion mass depends on its decay constant \cite{burgess2024uv}.

The expected presence of such a 4-form field $H$ in the low-energy theory endows the $B$ field with a nonzero mass. We verify this by introducing the Lagrange multiplier $a$ in the standard manner and integrating out the field $G$. This procedure yields the result:
\begin{equation} \label{L2Ca}
\mathcal{L}_2(C,a) =  -\frac{1}{2}(\partial a)^2 -J^{\mu} \partial_{\mu}a -\frac{1}{2}J_{\mu} J^{\mu} + \frac{1}{4!} (\mu_a a -\bar{\theta} \tilde{\Lambda}^2_{\text{QCD}}) \epsilon^{\mu\nu\lambda\rho} H_{\mu\nu\lambda\rho} -\frac{1}{2 \cdot 4!} H_{\mu\nu\lambda\rho} H^{\mu\nu\lambda\rho} + \cdot \cdot \cdot.
\end{equation}
Subsequent integration over $H$ leads to its saddle-point value, $H_{\mu\nu\lambda\rho} = \mathcal{H}_{\mu\nu\lambda\rho}$, where
\begin{equation} \label{Hepsilon}
\mathcal{H}_{\mu\nu\lambda\rho} = (\mu_a a -\bar{\theta} \tilde{\Lambda}^2_{\text{QCD}}) \epsilon_{\mu\nu\lambda\rho}.
\end{equation}
Substituting this solution back gives the effective axion Lagrangian:
\begin{equation} \label{L2a}
\mathcal{L}_2(a) =  -\frac{1}{2}(\partial a)^2 -J^{\mu} \partial_{\mu}a -\frac{1}{2}J_{\mu} J^{\mu} - \frac{1}{2} (\mu_a a -\bar{\theta} \tilde{\Lambda}^2_{\text{QCD}})^2,
\end{equation}
which demonstrates that the potential minimum indeed occurs at $a = \bar{\theta} \tilde{\Lambda}^2_{\text{QCD}}/\mu_a = \bar{\theta} f$. At this minimum, the CP-violating term in \eqref{L2Ca} vanishes. More details can be found in \cite{dvali2022strong, burgess2024uv, choi2023implications}.


\acknowledgments
Thanks for the discussion with Zhong-Zhi Xianyu, Haipeng An, Antonio Padilla and Paul Saffin. This work is supported by NSFC under Grants No. 12275146, the National Key R$\&$D Program of China (2021YFC2203100), the Dushi Program and the Shuimu Fellowship of Tsinghua University. For the purpose of open access, the authors have applied a CC BY public copyright licence to any Author Accepted Manuscript version arising.





\end{document}